\documentclass[11pt,a4paper]{article}
\usepackage{jheppub}
\usepackage{amsfonts,amsmath,hyperref,url, color}
\usepackage{eurosym}
\usepackage{braket}
\usepackage{graphicx}
\usepackage{mathrsfs}

\usepackage{xcolor}

\usepackage{amsmath,amsthm,amssymb}
\usepackage{physics}

\begin{document}

\date{\today}
\title{On symmetries and charges at spatial infinity}
\author[a]{Lennart Brocki}
\emailAdd{lennart.brocki@uwr.edu.pl}
\author[a,b]{Jerzy Kowalski-Glikman}
\affiliation[a]{University of Wroc\l{}aw, Faculty of Physics and Astronomy,  Maksa Borna 9, 50-204 Wroc\l{}aw, Poland}
\affiliation[b]{National Centre for Nuclear Research, Pasteura 7, 02-093 Warsaw, Poland}
\emailAdd{jerzy.kowalski-glikman@uwr.edu.pl}

\abstract{Following the recent work of Henneaux and Troessaert \cite{Henneaux:2018cst}, which revisits the problem of spacetime symmetries at spatial infinity, we analyze this problem using a Bondi-type metric without determinant condition as our starting point. It turns out  that in this case the symmetries at spatial infinity form  the BMS symmetry appended with an additional infinite set of abelian symmetries. We furthermore find that additionally imposing the determinant condition would result in a drastic reduction of symmetries, with no spatial (super) translations present.}


\maketitle

\section{Introduction}

Take a black box. There is some stuff in it and you want to know how to characterize its motion not looking inside the box. This can be done by measuring conserved charges given by integrals of specific functions of gravitational field. These charges are associated with physical symmetries, which in turn result from gauge symmetries of gravity being broken by the presence of boundaries (see \cite{Freidel:2020xyx} and references therein for in-depth discussion.) A particularly interesting physical situation arises when the boundary is situated far away from  localized material sources, where the geometry is almost flat. In the seminal paper \cite{Regge:1974zd}  Regge and Teitelboim formulated the well-defined variational principle for gravity on asymptotically flat space by adding an appropriate boundary term to the Einstein-Hilbert action and then show that the resulting boundary charges satisfy the Poincar\'e algebra and therefore measure the energy, momenta, angular momenta and boost charges of spacetime. This result was well expected and not surprising because the Poincar\'e symmetry is the physical symmetry of flat Minkowski space. A really surprising result was derived more than a decade before the work of Regge and Teitelboim, in \cite{Bondi:1962px}, \cite{Sachs:1962wk}, \cite{Sachs:1962zza} where Bondi, van der Burg, Metzner and Sachs established that the physical symmetries at null infinity of asymptotically flat spacetime are much larger and form an infinite dimensional group, the BMS group, containing Poincar\'e group as its (maximal) finite-dimensional subgroup (see \cite{Strominger:2017zoo} for a review of exciting  recent developments.) Then an immediate question arises, why the symmetries at null infinity are so much different from the ones at the spatial infinity? Or are they?

In the recent publication \cite{Henneaux:2018cst} Henneaux and Troessaert revisit the analysis \cite{Regge:1974zd} of Regge and Teitelboim of asymptotically flat spaces in the Hamiltonian (ADM) formulation and derived an infinite set of  charges generating BMS supertranslations at spatial infinity. Therefore the asymptotic symmetry of asymptotically flat spacetimes, according to \cite{Henneaux:2018cst}, is the (unextended) BMS group. This is in contrast to the original result of \cite{Regge:1974zd} where these charges were found to vanish and the asymptotic symmetry was determined to form the Poincar\'e group and not the BMS group. The key difference of the two approaches is that Henneaux and Troessaert make the asymptotic expansion in spherical coordinates that makes it possible for them to use parity conditions on the phase space functions different from the ones used by  Regge and Teitelboim in their case of Cartesian coordinates expansion. These conditions are the crucial point of the analysis because they guarantee cancellation of divergences, generally plaguing expressions for the asymptotic symplectic form and the charges.

The treatments in \cite{Henneaux:2018cst} and \cite{Regge:1974zd} have in common that they take as a starting point a generic asymptotic expansion of a spatial metric and its conjugate momenta. In this paper we instead perform a 3+1 decomposition of a spacetime metric $g$ in Bondi coordinates which is asymptotically flat at null infinity, off-shell and fulfills the Bondi gauge except for the determinant condition. We drop the determinant condition since, as it will be explained in detail below, its presence drastically reduces the asymptotic symmetries. Using the ADM formalism we express the expansion of the spatial metric and momenta in terms of the metric functions and their derivatives. These expressions are then substituted into the symplectic form, Hamiltonian and diffeomorphism constraint and charges given in \cite{Henneaux:2018cst}. This procedure leads to two interesting insights.

First, it shows that the falloff conditions on the momenta translate to conditions on retarded time derivatives of spacetime metric functions which describe the rate of gravitational radiation. If only spacetimes are allowed that radiate a finite amount of energy the falloff conditions are satisfied and we furthermore find that for the class of spacetimes and foliations we consider the asymptotic symplectic form term is finite without having to introduce parity conditions.

Second, we find an asymptotic symmetry at spatial infinity that is larger than the BMS symmetry. The crucial difference to the treatment of \cite{Henneaux:2018cst} is that we do not impose parity conditions such that a larger group of supertranslations leaves the boundary conditions invariant. The associated charges are shown to be finite. This larger-than-BMS algebra is isomorphic to the one found by Troessaert \cite{Troessaert:2017jcm}.

The plan of the paper is as follows. In the next section we recall the results presented in \cite{Henneaux:2018cst} and \cite{Regge:1974zd}.  In the following Sect.\ \ref{sec:momenta} we compute the conjugate momenta and express them in terms of the spacetime metric components. We then closely follow the treatment of \cite{Henneaux:2018cst} and analyze which differences arise when substituting these expressions for the momenta. Sect.\ \ref{sec:sympstruc} is devoted to the discussion of the asymptotic symplectic structure, while Sect.\ \ref{sec:constraint} concerns the Hamiltonian and diffeomorphism constraint. In Sect. \ \ref{sec:asympsymm} we discuss the consequences for the asymptotic symmetry. We end this paper with some concluding remarks in Sect.\ \ref{sec:conclusions}.

\section{Review of previous results }

In this section we recall some results derived in \cite{Henneaux:2018cst} and \cite{Regge:1974zd} that we are going to make use of in the paper.

\subsection{Role of surface integrals and extension of the original treatment to include BMS}\label{sec:role}
Since it is crucial for our discussion, we briefly review the Hamiltonian analysis made in \cite{Regge:1974zd}.
The main result is that in order to have a well-defined Hamiltonian formalism one has to add surface integrals to the Hamiltonian which, for asymptotically flat spacetimes, turn out to be the Poincar\'e charges. Asymptotically flat spacetimes and the associated conjugate momenta in asymptotically Cartesian coordinates are assumed to obey the asymptotic expansion ($r^2 =\sum_i x_i^2$)
\begin{align}
	g_{ij}&=\delta_{ij}+\frac{1}{r}\bar h_{ij}+O(r^2)\label{asymcart1}\\
	\pi^{ij}&=\frac{1}{r^2}\bar \pi^{ij}+O(r^3).\label{asymcart2}
\end{align}
Hamilton's equations are defined as the functional derivatives
\begin{align}
	\dot g_{ij} &= \frac{\delta H}{\delta\pi^{ij}}\\
	\dot \pi^{ij}&=-\frac{\delta H}{\delta g_{ij}}
\end{align}
which are, by definition, the coefficients of $\delta g_{ij}$ and $\delta\pi^{ij}$ in the variation of the Hamiltonian
\begin{equation}\label{varH}
	\delta H = \int d^3x[A^{ij}\delta g_{ij}+B_{ij}\delta\pi^{ij}].
\end{equation}
Therefore, for Hamilton's equations to be defined properly, it is necessary that the variation of the Hamiltonian can be put into the form of \eqref{varH}. The important observation made in \cite{Regge:1974zd} is that this is not the case for the Hamiltonian constraint of general relativity
\begin{equation}
	H_0 = \int d^3x N\mathcal H + N^i\mathcal H_i
\end{equation}
but that instead extra surface terms appear
\begin{align}\label{surfaceterm}
	\delta H_0 = \int d^3x[A^{ij}\delta g_{ij}+B_{ij}\delta\pi^{ij}] + \int d^2x K_{N,N^i},
\end{align}
which arise as a result of  partial integration  moving a derivative away from the variations of the canonical variables.
If the boundary term in the last expression can be rewritten as the variation of some quantity $C$, i.e.\ if it is exact
\begin{equation}\label{varC}
\delta H_0 = \int d^3x[A^{ij}\delta g_{ij}+B_{ij}\delta\pi^{ij}] - \delta C_{N,N^i}
\end{equation}
this quantity can now be added to the Hamiltonian, such that the variation of the redefined Hamiltonian is of the form
\begin{align}
	\delta(H_0+C_{N,N^i}) =  \int d^3x[A^{ij}\delta g_{ij}+B_{ij}\delta\pi^{ij}]
\end{align}
and thus the functional derivatives are well defined. Then if the surface integral $C$ is finite, it turns out to be a combination of Poincar\'e charges that satisfy the Poincar\'e algebra. Furthermore the transformations generated by these charges leave the asymptotic conditions \eqref{asymcart1}, \eqref{asymcart2} invariant. Physically, this means that the asymptotic observer can measure the total momentum and angular momentum, as well as the boost, of asymptotically flat spacetime. In order to make the charges and asymptotic symplectic structure  finite Regge and Teitelboim introduce certain parity conditions on the leading order terms $\bar h_{ij}$ and $\bar\pi^{ij}$ in the metric expansion \cite{Regge:1974zd}, where parity here refers to antipodal points on the sphere, i.e. the map $\mathbf{n} \rightarrow - \mathbf{n}$, where $\mathbf{n}$ is a unit vector. Explicitly, it is  assumed that
\begin{equation}\label{RTparity}
	\bar h_{ij}(-\mathbf{n})=\bar h_{ij}(\mathbf{n})\,,\quad \bar \pi_{ij}(-\mathbf{n})=-\bar \pi_{ij}(\mathbf{n})\,.
\end{equation}
so that $\bar h_{ij}$ is even under parity, while $\bar\pi^{ij}$ is odd.

It is crucial to notice that approach of Regge and Teitelboim in addition to the standard Poincar\'e translations and Lorentz transformations the asymptotic conditions are also invariant under angle-dependent translations, however, the associated charges vanish due to the introduced parity conditions. This means that the actual asymptotic symmetry algebra is the Poincar\'e algebra since the symmetry under angle-dependent translations is pure gauge, i.e., the charges associated with these symmetries vanish as a consequence of the chosen parity conditions\footnote{We will discuss the relation between gauge symmetries, boundary charges and asymptotic conditions in more details below.}.

In the recent paper \cite{Henneaux:2018cst} Henneaux and Troessaert  propose different parity conditions which keep Hamilton's equations well-defined but are less restrictive in the sense that they render the charges associated with angle-dependent translations finite. They further show that the algebra of these charges is isomorphic to the BMS algebra.

Instead of Cartesian coordinates used by Regge and Teitelboim, in \cite{Henneaux:2018cst} Henneaux and Troessaert employ spherical coordinates and the asymptotic conditions take the form
\begin{align}\label{falloff}
	h_{rr} &= 1+\frac 1r \bar h_{rr} + \frac{1}{r^2}h_{rr}^{(2)}+o(r^{-2}) \\
	h_{rA} &= \bar h_{rA}+\frac{1}{r}h_{rA}^{(2)}+o(r^{-1})\label{falloffrA}\\
	h_{AB} &= r^2\bar{\gamma}_{AB} + r\bar h_{AB}+h_{AB}^{(2)}+o(1) \label{falloff3}\\
	\pi^{rr} &= \bar\pi^{rr} + \frac{1}{r}\pi^{(2)rr}+o(r^{-1})\label{falloffrr}\\
	\pi^{rA} &= \frac{1}{r}\bar\pi^{rA}+\frac{1}{r^2}\pi^{(2)rA}+o(r^{-2})\\
	\pi^{AB} &= \frac{1}{r^2}\bar{\pi}^{AB} +\frac{1}{r^3}\pi^{(2)AB}+ o(r^{-3})\label{falloff1}
\end{align}
while the parity conditions are
\begin{equation}\label{parity}
	\bar{\lambda} \sim \bar{\pi}^{A B}=\operatorname{even}, \quad \bar{p} \sim \bar{k}_{A B} \sim \bar{\pi}^{r A}=\mathrm{odd},
\end{equation}
where
\begin{align}
		\bar{\lambda}&=\frac{1}{2} \bar h_{r r}, \quad \bar{k}_{A B}=\frac{1}{2} \bar{h}_{A B}+\bar{\lambda} {\gamma}_{A B}\label{defkAB} \\
		\bar{p}&=2\left(\bar{\pi}^{r r}-\bar{\pi}_{A}^{A}\right), \quad \pi_{(k)}^{A B}=2 \bar{\pi}^{A B}.\label{2.18}
\end{align}
and $\gamma_{AB}$ is the metric on the unit sphere. 
Although a generic expansion of an asymptotically flat metric allows the term $\bar h_{rA}$ to be non-vanishing Henneaux and Troessaert assume \cite{Henneaux:2018cst} that
\begin{align}\label{hravanish}
	\bar h_{rA}=0
\end{align}
which is necessary in order for the boost charges to be integrable.

The parity conditions are introduced to cancel the following logarithmic divergences in the Hamiltonian kinetic term, i.e., the symplectic structure
\begin{equation}\label{sympstr}
	\int dr\frac 1r\int d \theta d \varphi\left(\bar{\pi}^{r r} \dot{\bar{h}}_{r r}+\bar{\pi}^{A B} \dot{\bar{h}}_{A B}\right)=\int dr\frac 1r\int d \theta d \varphi\left(\bar{p} \dot{\bar{\lambda}}+\pi_{(k)}^{A B} \dot{\bar{k}}_{A B}\right),
\end{equation}
which is zero because the integral over the sphere of a function with odd parity vanishes. It is furthermore demonstrated that all divergences occurring in the expressions for the charges can be canceled by imposing   the Hamiltonian and diffeomorphism constraints to the leading order and  that  no parity conditions have to be imposed. The vanishing of the leading order of constraints imposes the following relations
\begin{align}
	\bar \pi^{rA}&=-D_B\bar \pi^{BA}\label{constraintA}\\
	D_A\bar \pi^{Ar}&=\bar \pi^A_A\label{constraintr}\\
	D_AD_B \bar k^{AB}&=D_AD^A \bar k,\label{constraint},
\end{align}
which arise from $\mathcal H^A, \mathcal H^r$ and $\mathcal H$ respectively.

The transformations preserving the above boundary conditions are
\begin{align}\label{transfos}
		\xi^{\perp}=r b+F+O\left(r^{-1}\right), \quad \xi^{r}=W+O\left(r^{-1}\right), \quad \xi^{A}=Y^{A}+\frac{1}{r} I^{A}+O\left(r^{-2}\right) 
\end{align}
with
\begin{equation}\label{transfoscond}
{D}_{A} {D}_{B} b+{\gamma}_{A B} b=0, \quad \mathcal{L}_{Y} {\gamma}_{A B}=0,
\end{equation}
where $b, F, W, Y^A$ are functions on the sphere and
\begin{equation}
I^A=\frac{2b}{\sqrt{\gamma}}\bar\pi^{rA}+D^A W
\end{equation}

The vectors $Y^A$ describe spatial rotations, $b$ Lorentz boosts, $f$ contains time translations through its zero mode and $W$ contains spatial translations through the $l=1$ terms in an expansion in spherical harmonics. In order for the parity conditions \eqref{parity} to be preserved as well Henneaux and Troessaert  further assume \cite{Henneaux:2018cst}
\begin{equation}\label{defF}
F=-3b\bar \lambda -\frac 12 b\bar h+T,
\end{equation}
where $T$ is an even function on the sphere and $W$ is an odd function. The above defined transformations $\xi$ form under the bracket \cite{Brown:1986nw}
\begin{equation}\label{defbracket}
	\left[\xi_{1}, \xi_{2}\right]_{M}^{\mu}=\left[\xi_{1}, \xi_{2}\right]_{S D}^{\mu}+\delta_{2}^{h, \pi} \xi_{1}^{\mu}-\delta_{1}^{h, \pi} \xi_{2}^{\mu}-\oint d^2x \{\xi_1^{\mu},\xi_2^{\nu}\}\mathcal H_{\nu},
\end{equation}
where $\mathcal H_{\nu}=(\mathcal H, \mathcal H_i)$ and $\{\xi_1^{\mu},\xi_2^{\nu}\}$ is the Poisson bracket, the following algebra
\begin{equation}\label{algatspatial}
    \widehat{\xi}(\widehat{Y}, \widehat{b}, \widehat{T}, \widehat{W})=\left[\xi_{1}\left(Y_{1}, b_{1}, T_{1}, W_{1}\right), \xi_{2}\left(Y_{2}, b_{2}, T_{2}, W_{2}\right)\right]_{M},
\end{equation}
with
\begin{align}
	\widehat{Y}^{A} &=Y_{1}^{B} \partial_{B} Y_{2}^{A}+\bar{\gamma}^{A B} b_{1} \partial_{B} b_{2}-(1 \leftrightarrow 2) \\
	\widehat{b} &=Y_{1}^{B} \partial_{B} b_{2}-(1 \leftrightarrow 2) \\
	\widehat{T} &=Y_{1}^{A} \partial_{A} T_{2}-3 b_{1} W_{2}-\partial_{A} b_{1} {D}^{A} W_{2}-b_{1} {D}_{A} {D}^{A} W_{2}-(1 \leftrightarrow 2) \\
	\widehat{W} &=Y_{1}^{A} \partial_{A} W_{2}-b_{1} T_{2}-(1 \leftrightarrow 2),
\end{align}
which is shown to be isomorphic to the BMS algebra. In \eqref{defbracket}   
$\left[\xi_{1}, \xi_{2}\right]_{S D}$ is the surface deformation bracket \cite{Teitelboim:1972vw} defined as
\begin{align}
    \left[\xi_{1}, \xi_{2}\right]^{\perp}_{S D}&=\xi_1^i\partial_i\xi_2^{\perp}-\xi_2^i\partial_i\xi_1^{\perp}\\
    \left[\xi_{1}, \xi_{2}\right]^i_{S D}&=\xi_1^k\partial_k\xi_2^i-\xi_2^k\partial_k\xi_1^i+h^{ik}(\xi_1^{\perp}\partial_k\xi_2^{\perp}-\xi_2^{\perp}\partial_k\xi_1^{\perp})
\end{align}
and $\delta_2^{h,\pi}\xi_1$ is given by
\begin{align}
     \delta_2^{h,\pi}\xi_1=\frac{\delta \xi_1}{\delta g_{ij}}\delta_{\xi_2}  g_{ij}+\frac{\delta \xi_1}{\delta \pi^{ij}}\delta_{\xi_2}  \pi^{ij}.
\end{align}
The terms of the form $\delta_{1}^{g, \pi} \xi_{2}$ appear because $\xi$ depends on phase-space functions and one therefore has to take into account the change of $\xi$ induced by the variation of these functions. Recently, the meaning of the surface deformation bracket was clarified in \cite{Ciambelli:2021nmv} and \cite{Freidel:2021dxw}.

The boundary terms given in \cite{Henneaux:2018cst}, which correspond to the term $\int d^2x K$ in \eqref{surfaceterm},  are
\begin{equation}\label{surfaceterms}
\begin{aligned}
\mathcal{K}_{\xi}\left[\delta g_{i j}, \delta \pi^{i j}\right]=\delta \oint d^{2} x &\left\{-2 Y^{A}\left(\bar{h}_{A B} \bar{\pi}^{r B}+\bar{\gamma}_{A B} \pi^{(2) r B}+\bar h_{rA} \bar{\pi}^{r r}\right)\right.\\
&\left.-2 \sqrt{\gamma} b k^{(2)}-\sqrt{\bar{\gamma}} \frac{1}{4} b\left(\bar{h}^{2}+\bar{h}^{A B} \bar{h}_{A B}\right)\right\} \\
+\oint d^{2} x &\left\{-2 I^{A} \bar{\gamma}_{A B} \delta \bar{\pi}^{r B}-2 W \delta \bar{\pi}^{r r}-\sqrt{\bar{\gamma}}(2 F+\bar{h} b) \delta\left(2 \bar{\lambda}+\bar{D}_{A} \bar h_{rA}\right)\right.\\
&\left.+\sqrt{\bar{\gamma}}\left(\bar h^{rC}\partial_{C} b \bar{\gamma}^{A B}-b \bar{D}^{A} \bar h^{rB}\right) \delta \bar{h}_{A B}\right\}+o\left(r^{0}\right).
\end{aligned}
\end{equation}
It can be seen that the form of $F$ in \eqref{defF} guarantees integrability of the boundary terms. The expression for the charge, which corresponds to the term $C$ in \eqref{varC}, is finally given by
\begin{equation}\label{charge1}
\begin{aligned}
\mathcal{B}_{\xi}\left[g_{i j}, \pi^{i j}\right]=& \oint d^{2} x\left\{Y^{A}\left(4 \bar{k}_{A B} \bar{\pi}^{r B}-4 \bar{\lambda} \bar{\gamma}_{A B} \bar{\pi}^{r B}+2 \bar{\gamma}_{A B} \pi^{(2) r B}\right)+W\left(2 \bar{\pi}^{r r}-2 \bar{D}_{A} \bar{\pi}^{r A}\right)\right.\\
&\left.+T 4 \sqrt{\bar{\gamma}} \bar{\lambda}+b \sqrt{\bar{\gamma}}\left(2 k^{(2)}+\bar{k}^{2}+\bar{k}_{B}^{A} \bar{k}_{A}^{B}-6 \overline{\lambda k}\right)+b \frac{2}{\sqrt{\gamma}} \bar{\gamma}_{A B} \bar{\pi}^{r A} \bar{\pi}^{r B}\right\}
\end{aligned}
\end{equation}
and notice in particular that the charges proportional to higher modes of $W$ and $T$ are in general non-vanishing. Supertranslations are therefore part of the asymptotic symmetry for the new set of parity conditions.

To see how the cancellation of the divergences in the boundary terms works consider for instance the divergence proportional to $Y^A$, which appears in the term $\int d^2x K$ in \eqref{surfaceterm} (see \cite{Henneaux:2018cst} for details of derivation of the charges)
\begin{align}
	\int d^2x K_Y&=-2r\oint d^2x Y^A\gamma_{AB}\delta\bar{\pi}^{rB} + O(1)\nonumber\\
	&=2r\oint d^2x Y_BD_C\delta \bar \pi^{BC}+O(1)\nonumber\\
	&=-2r\oint d^2x D_{(C}Y_{B)}\delta \bar \pi^{BC}+O(1),\nonumber
\end{align}
which vanishes since $Y^A$ are the Killing vectors on the sphere and thus obey 
\begin{equation}
	D_{(C}Y_{B)}=0\label{YKilling}.
\end{equation}
This means that in order to remove potential $Y$-charges divergences we must assume that we cannot extend the rotation sector beyond the standard three rotational Killing vectors.

The aim of this paper is to express, in a first step, the expansions of the spatial metric and conjugate momenta \eqref{falloff}-\eqref{falloff1} in terms of an asymptotically flat spacetime metric. This will be done by means of a 3+1 decomposition and the ADM formalism \cite{Arnowitt:1962hi}, which we briefly review in the Appendix A. In a second step we substitute these expressions into the symplectic structure \eqref{sympstr}, constraints \eqref{constraintA}-\eqref{constraint} and charges \eqref{charge1} and analyze what are the consequences of this procedure for the asymptotic symmetries.

\section{Conjugate momenta in terms of spacetime metric components}\label{sec:momenta}
In this section we are going to perform a 3+1 decomposition of a spacetime metric and use the ADM formalism to express the momenta in terms of components of this metric and their derivatives. 
We consider a metric that is asymptotically flat at null infinity and is defined by the following expansion
\begin{align}\label{asympflat}
	g_{\mu\nu}dx^{\mu}dx^{\nu} &= -\left(1-\frac{2M}{r}-\frac{\bar g_{uu}}{r^2}+O(r^{-3})\right)du^2-2\left(1-\frac{\bar g_{ur}}{r}-\frac{g_{ur}^{(2)}}{r^2}+O(r^{-3})\right)dudr\nonumber\\
	&+\left(\psi_A+\frac{1}{r}F_A+O(r^{-2})\right)dudx^A\nonumber\\
	&+\left(r^2\gamma_{AB}+rC_{AB}+D_{AB}+O(r^{-1})\right)dx^Adx^B,
\end{align}
where $\gamma_{AB}$ is the unit metric on the sphere and all other metric components are functions of $(u,x^A)$. This metric is subject only to the partial Bondi gauge condition
\begin{align}\label{partialBondi}
	g_{rr}=0,\quad g_{rA}=0
\end{align}
and no further assumptions are made at this stage. 

The metric \eqref{asympflat} is more general than the Bondi metric \cite{Bondi:1962px,Sachs:1962wk,Sachs:1962zza}, \cite{Flanagan:2015pxa}, \cite{Barnich:2009se}, which usually is assumed to additionally satisfy the Einstein field equations and to be subject to the determinant condition 
\begin{align}\label{detcond}
    \det g_{AB}=r^4\det \gamma_{AB},
\end{align}
which implies
\begin{align}
    \gamma^{AB}C_{AB}=0,\quad \gamma^{AB}D_{AB}=\frac 12 C^{AB}C_{AB}.
\end{align}
Here we instead start with an off-shell metric and the field equations will be partially imposed by demanding that the leading order of the Hamiltonian and diffeomorphism constraint has to vanish, as was done in \cite{Henneaux:2018cst}. Furthermore, we are not imposing the determinant condition since it leads to a metric that is too rigid: in subsection \ref{subsec:detcond} we will demonstrate that imposing this condition would drastically reduce the asymptotic symmetry by excluding spatial (super) translations.

We choose spacelike hypersurfaces $\Sigma_t$ of constant `time' $t$ defined by
\begin{align}\label{deft}
    t=u+r+f(x^A)+\frac{g(x^A)}{r}    
\end{align}
and in coordinates $(t,r,x^A)$ the metric \eqref{asympflat} takes the form
\begin{align}\label{metrict}
	g_{\mu\nu}dx^{\mu}dx^{\nu} &= -\left(1-\frac{2M}{r}-\frac{\bar g_{uu}}{r^2}+O(r^{-3})\right)dt^2+2\left(\frac{\bar g_{ur}-2M}{r}+O(r^{-2})\right)dtdr\nonumber\\
	&+2\left(\partial_A f+\frac 12 \psi_A+O(r^{-1})\right)dtdx^A\nonumber\\
	&+\left(1+\frac{2M-2\bar g_{ur}}{r}+\frac{\bar g_{uu}-2 g_{ur}^{(2)}}{r^2}+O(r^{-3})\right)dr^2\nonumber\\
	&+\left(-\psi_A+\frac{(4M-2\bar g_{ur})\partial_A f-F_A}{2r}+O(r^{-2})\right)drdx^A\nonumber\\
	&+\left(r^2\gamma_{AB}+rC_{AB}+D_{AB}-\partial_Af\partial_Bf-\psi_A\partial_Bf\right)dx^Adx^B.
\end{align}
The foliation defined by \eqref{deft} is chosen such that the resulting induced metric $h_{ab}$, as defined by \eqref{decomp}, agrees with the fall-off conditions \eqref{falloff}-\eqref{falloff3}. This rules out the presence of a term $r k(x^A)$ in $t$ since it would lead to 
\begin{equation}
	h_{ab}dy^ady^b=dr^2(2k-k^2+O(r^{-1}))+..,
\end{equation}
which does not agree with \eqref{falloff}.
It also rules out a logarithmic term $\log(r)h(x^A)$ since it would lead to a term proportional to $\log(r)/r\partial_A hdrdx^A$ to be present in $h_{ab}$, which is not allowed by \eqref{falloffrA}. Notice that our choice of $t$ therefore in particular excludes Schwarzschild time defined by $t'=u+r+2M\log(r/2M-1)$, with $M=const.$, this lack is however unproblematic in the sense that, as we are going to show, we obtain the correct expression for the ADM mass of the Schwarzschild metric, which is a special case of \eqref{asympflat} with $M=const$ and all subleading components vanishing. This result is in line with the expectation that the choice of foliation is arbitrary and does not affect the charges, as long as the resulting normal vector on $\Sigma_t$ is timelike and $h_{ab}$ is asymptotically flat.

By comparing the form of \eqref{metrict} with the decomposition \eqref{decomp} we find  the following expressions for lapse and shift
\begin{align}\label{lapseshift}
	N=1-\frac Mr+O(r^{-2}),\quad N_r=\frac{\bar g_{ur} - 2M}{r}+O(r^{-2}),\quad N_A=\partial_Af+\frac 12 \psi_A+O(r^{-1})
\end{align}
and we identify the leading order terms in the metric expansion \eqref{falloff}-\eqref{falloff3} as
\begin{align}\label{identify}
	&\bar h_{rr}=2M-2\bar g_{ur},\quad \bar h_{rA}=-\frac{\psi_A}{2},\quad \bar h_{AB}=C_{AB}\nonumber\\
	& h_{rr}^{(2)}=\bar g_{uu}-2g_{ur}^{(2)},\quad h_{rA}^{(2)}=\frac{4\partial_A f-F_A}{2},\quad h_{AB}^{(2)}=D_{AB}-\partial_Af\partial_Bf-\psi_A\partial_Bf.
\end{align}
{As they were defined in \eqref{asympflat} the spacetime metric functions like $M$ are functions of $(u, x^A)$, which might seem a bit odd since they now appear in the components of the induced metric $h_{ab}$ which is described by components $(r, x^A)$. But owing to \eqref{deft} $u$ is not an independent coordinate on a hypersurface $\Sigma_t$ and neither is $t$ which is now understood as a parameter labeling these surfaces, such that $h_{ab}$ can be expressed purely in terms of $(r, x^A)$. In particular this implies that on a surface $\Sigma_t$ in the large $r$ limit, in which the expressions \eqref{identify} are defined, we have $u \rightarrow -\infty$, which is the defining limit for spatial infinity $i^0$. This limit for $u$ is from now on implied throughout whenever the spacetime metric functions appear.}
The condition \eqref{hravanish} now takes the form
\begin{align}\label{vanishpsi}
	\psi_A=0,
\end{align}
and following the arguments of \cite{Henneaux:2018cst} we will eventually also make this assumption. For the sake of obtaining a general form for the expressions of the momenta, however, we will keep $\psi_A$ non-zero for now and assume that it is vanishing from Sect.\ \ref{sec:constraint} on. 

The unit normal on the spacelike hypersurfaces is given by
\begin{equation}
	n_{\alpha}=-N\partial_{\alpha} t=-\left(1-\frac Mr+O(r^{-2})\right)\partial_{\alpha}\left(u+r+f(x^A)+\frac{g(x^A)}{r}\right)
\end{equation}
and using this expression to evaluate the extrinsic curvature \eqref{extrinsic} we find the asymptotic expressions  
\begin{align}\label{extrinsic1}
	K_{rr}&=-\frac{\partial_u M}{r}+O(r^{-2}),\\
	K_{rA}&=-\frac{1}{4r}\Big(-2\partial_A \bar g_{ur}+4\partial_A M-2\left(\psi_A+2\partial_A f\right)\nonumber\\
	&\quad\quad-(\psi_B+2\partial_B f)\gamma^{BC}\partial_u C_{CA}-4\partial_A f\partial_u M\Big)+O(r^{-2})\label{extrinsicrA}\\
	K_{AB}&=\frac r2 \partial_u C_{AB}+O(r^{-2}).\label{extrinsic2}
\end{align}
We have now all the expressions at hand we need to write the momenta in terms of the components of \eqref{asympflat} and find from evaluating \eqref{momenta}
\begin{align}\label{momentumrr}
	\pi^{rr}&=-\frac{r}{2}\sqrt{\gamma}\gamma^{AB}\partial_uC_{AB}+\frac{1}{2}\sqrt{\gamma}\big(4\mathcal M-\gamma^{AB}\partial_u D_{AB}\nonumber\\
	&\quad +\gamma^{AB}D_A(2D_B f+\psi_B)+G[\partial_uC_{AB}]\big)+O(r^{-1}) \\
	\pi^{rA}&=\frac{1}{4r}\sqrt{\gamma}\left(-2\gamma^{AB}D_B\mathcal M+2\gamma^{AB}(2D_B f+\psi_B)+G[\partial_u M, \partial_u C_{AB}]\right)+O(r^{-2})\label{momentumrA}\\
	\pi^{AB}&=\frac{1}{2r}\sqrt{\gamma}\left(2\gamma^{AB}\partial_u M-\left(\gamma^{AB}\gamma^{CD}\partial_uC_{CD}-\partial_u C^{AB}\right)\right)\nonumber\\
	&+\frac{1}{2r^2}\sqrt{\gamma}\Big[\gamma^{AB}\left(\partial_u\bar g_{uu}-2\mathcal M\partial_u \bar g_{ur}\right)-\left(\gamma^{AB}\gamma^{CD}\partial_uD_{CD}-\partial_u D^{AB}\right)\nonumber\\
	&\quad\quad-\frac{\gamma^{AB}}{2}(\gamma^{CD}(2D_C f+\psi_C)\partial_u \psi_D)-(2D^{A}D^{B}f+D^{(A}\psi^{B)})\nonumber\\
    &\quad\quad	+\gamma^{CD}(2D_CD_D f+D_C\psi_D)\gamma^{AB}+G[\partial_u M, \partial_u C_{AB}]
	\Big]+O(r^{-3}),\label{momentumAB}
\end{align}
where 
\begin{equation}
	\mathcal M=\bar g_{ur}-2M
\end{equation} 
and $G$ stands for lengthy terms proportional to either $\partial_u C_{AB}$ or $\partial_u M$. It might seem strange that derivatives over $u$ appear in the above expressions for the extrinsic curvature and conjugate momenta since they live on a spacelike surface with coordinates $(r,x^A)$. But using \eqref{deft} we could write $\partial_u$ in terms of derivatives over $(t, r, x^A)$ and setting in the resulting expressions $t=const.$ we find that the curvature and momenta are described solely in terms of $(r, x^A)$, as they should. 

{Comparing the expressions \eqref{momentumrr}-\eqref{momentumAB} with the falloff conditions \eqref{falloffrr}-\eqref{falloff1} we see that the $O(r)$ term of $\pi^{rr}$ and the $O(r^{-1})$ term of $\pi^{AB}$ should be vanishing. Thus we find that the falloff conditions translate into the conditions on the terms $\partial_u C_{AB}$ and $\partial_u M$ near spatial infinity, i.e. in the limit $u \rightarrow -\infty$. And recalling the fact that on-shell these terms describe the rate of gravitational radiation, as was explained in \cite{Bondi:1962px} (see also \cite{Compere:2018aar}), we find that it is a natural physical requirement that near spatial infinity these derivatives over $u$ behave as
\begin{align}\label{damping}
	\partial_u C_{AB}\sim u^{-(1+\varepsilon)},\quad \partial_u M \sim u^{-(1+\varepsilon)},\quad \varepsilon > 0,
\end{align}
since otherwise the amount of radiated energy would be divergent. Written in terms of the coordinates on the spacelike slice $(r,x^A)$ these conditions read
\begin{align}\label{damping1}
    \partial_uC_{AB}=\frac{\hat C_{AB}(x^A)}{r^{1+\epsilon}},\quad \partial_u M=\frac{\hat M(x^A)}{r^{1+\epsilon}},
\end{align}
where $\hat M$ and $\hat C_{AB}$ are arbitrary functions on the sphere. They therefore provide an extra damping factor so that the $O(r)$ contribution of $\bar \pi^{rr}$ and $O(r^{-1})$ of $\bar \pi^{AB}$ are vanishing and we will only consider such spacetimes that satisfy these conditions. Notice that also the terms denoted by $G$ in the expressions for the momenta are vanishing under the condition \eqref{damping1}. The requirement to only allow such ``physically reasonable" spacetimes has already been pointed out in \cite{Regge:1974zd}.}

\section{Finiteness of symplectic structure}\label{sec:sympstruc}
The falloff conditions defined in \eqref{falloff}-\eqref{falloff1} are not sufficient to remove divergences in the symplectic structure
\begin{align}\label{sympstr1}
	\int d^3x \pi^{ab}\dot h_{ab},
\end{align}
since terms of order $O(r^{-1})$ appear which are logarithmically divergent.
As was explained above, after eq.\ \eqref{sympstr}, the authors of \cite{Henneaux:2018cst} remove these divergences by introducing parity conditions on the leading order terms in the expansion of the metric and momenta. The terms which are potentially divergent are the following ones
\begin{align}\label{kinetic}
	\int dr\frac 1r\int d\phi d\theta \left(\bar \pi^{rr} \dot{\bar h}_{rr}+\bar \pi^{AB} \dot{\bar h}_{AB}+\bar{\pi}^{rA}\dot{\bar h}_{rA}\right)
\end{align}
and \cite{Henneaux:2018cst} introduces parity conditions such that the integral over the sphere vanishes. 

Generically the components of $\dot {\bar h}_{ab}$ in \eqref{kinetic} are finite but they are in fact vanishing for the case that we consider. To show this we use \eqref{velocity} to express  $\dot {\bar h}_{ab}$ in terms of the momenta $\pi^{ab}$, lapse $N$ and shift $N_a$, whose asymptotic behavior is determined by the class of spacetimes we consider, defined by \eqref{asympflat} and \eqref{damping1}, and by the foliation, defined by \eqref{deft}. Evaluating \eqref{velocity} with the expressions for the momenta \eqref{momentumrr}-\eqref{momentumAB} and lapse and shift \eqref{lapseshift} we find
\begin{align}
	\dot {h}_{rr}&=-\frac{2\partial_u M}{r}+O(r^{-2}), \\
	\dot {h}_{rA}&=\frac{1}{r}\left(\bar K_{rA}-(2\partial_A f+\psi_A)+\partial_A\mathcal M\right)+O(r^{-2}),\\
	\dot {h}_{AB}&=r\partial_u C_{AB}+O(1),
\end{align}
where $\bar K_{rA}$ is the leading order of \eqref{extrinsicrA}. If we now use the condition \eqref{damping1} we find that $\dot{\bar h}_{ab}$ is indeed vanishing and so is \eqref{kinetic}. In the case that we consider there is therefore no need to impose parity conditions. This potentially enlarges the asymptotic symmetry, since the supertranslations do not have to be restricted to preserve the parity of the canonical fields, as was done in \cite{Henneaux:2018cst}. This enlargement of symmetry is physical only if the associated charges are non-vanishing, which we are going to check in the following sections.

\section{Leading order of constraints}\label{sec:constraint}
As it was explained in subsection \ref{sec:role} the vanishing of the leading order of the Hamiltonian \eqref{constraint} and diffeomorphism constraint \eqref{constraintA}, \eqref{constraintr} is crucial for canceling divergences which arise in the expression of the charges. In this section we show which restrictions on the form of the momenta it implies. We are also going to assume $\psi_A=0$ from now on.

Substituting \eqref{momentumrA} into \eqref{constraintr} we obtain
\begin{align}
	\frac{\sqrt{\gamma}}{2}\gamma^{AB}D_B(2f-\mathcal M)=-D_B\bar \pi^{AB}
\end{align}
which implies
\begin{align}\label{constrpiAB}
	\frac{\sqrt{\gamma}}{2}\gamma^{AB}(2f-\mathcal M)=-\bar \pi^{AB}+ A \gamma^{AB}\sqrt{\gamma}\;,
\end{align}
with $A$ being an arbitrary constant. Plugging in \eqref{momentumAB} this equation expresses a relation between several spacetime metric functions at spatial infinity 
\begin{align}\label{eqAB}
	\frac{\gamma^{AB}}{2}(2f-\mathcal M) = -&\gamma^{AB}(\partial_u \bar g_{uu}-2\mathcal M\partial_u \bar g_{ur}-\gamma^{CD}\partial_u D_{CD}+2D^2f)\nonumber\\
	&-\partial_uD^{AB}+2D^AD^Bf+A\gamma^{AB}\sqrt{\gamma}.
\end{align}
Solving \eqref{constrpiAB} for $\bar \pi^{AB}$ and substituting it together with \eqref{momentumrA} into \eqref{constraintr} yields
\begin{align}
	(D^2+2)(2f-\mathcal M)=4A
\end{align}
and upon expanding $2f-\mathcal M$ in spherical harmonics $Y_{lm}$ satisfying $D^2 Y_{lm} = -l(l+1) Y_{lm}$ we find that this equation has a general solution of the form
\begin{equation}\label{condfM}
	2f-\mathcal M = 2A + \sum_{m=-1}^1 a_m Y_{1m}
\end{equation}
with $a_m$ being arbitrary constants.

Finally, we have another condition coming from \eqref{constraint}, which reads
\begin{align}\label{condH}
    D^AD^BC_{AB}-D^2\gamma^{AB}C_{AB}=2D^2(M-\bar g_{ur}),
\end{align}
where we have used the definitions in \eqref{defkAB} and \eqref{identify}.
To summarize, the momenta are now expressed in terms of the spacetime metric functions as
\begin{align}
	\bar \pi^{rr}=&\frac{\sqrt{\gamma}}{2}(4\mathcal M+2D^2f-\gamma^{AB}\partial_u D_{AB})\label{pirr},\\
	\bar \pi^{rA}=&\frac{\sqrt{\gamma}}{2}\gamma^{AB}D_B(2f-\mathcal M)\label{pirA},\\
	\bar \pi^{AB}=&\frac{\sqrt{\gamma}}{2}\gamma^{AB}(\mathcal M-2f+2A),\label{piBA}
\end{align}
subject to the conditions \eqref{eqAB}, \eqref{condfM} and \eqref{condH}.

\section{Asymptotic symmetries}\label{sec:asympsymm}
In this section we discuss the asymptotic symmetries of $h_{ab}$ and $\pi^{ab}$, in particular we will analyze which transformations preserve the falloff conditions \eqref{falloff}-\eqref{falloff1} and the gauge condition \eqref{hravanish}. This will reproduce the expressions \eqref{transfos} and \eqref{transfoscond} which were already given in \cite{Henneaux:2018cst}. We are giving here the details of this derivation to stress the fact that, as we are going to show, the preservation of the falloff conditions allows for a large group of supertranslations parametrized by two arbitrary functions on the sphere. These details are furthermore needed for our discussion of the determinant condition \eqref{detcond} in subsection \ref{subsec:detcond}. 

To this end we will evaluate the change in the canonical variables generated by $G_{\xi}=\int d^3x(\xi^{\perp}\mathcal H+\xi^a\mathcal H_a)$ which is given by, see \cite{Arnowitt:1962hi} 
\begin{align}\label{transformationhab}
	\delta_{\xi} h_{ab}=2 \xi^{\perp} h^{-1 / 2}\left(\pi_{ab}-\frac{1}{2} h_{ab} \pi\right)+\mathcal{L}_{\xi} h_{ab}.
\end{align}

\begin{equation}\label{transformationpab}
	\begin{aligned}
		\delta \pi^{ab}=&-\xi^{\perp} h^{\frac{1}{2}}\left(R^{ab}-\frac{1}{2} h^{ab} R\right)+\frac{1}{2} \xi^{\perp} h^{ab} h^{-\frac{1}{2}}\left(\pi_{m n} \pi^{m n}-\frac{1}{2} \pi^{2}\right) \\
		&-2 \xi^{\perp} h^{-\frac{1}{2}}\left(\pi^{a m} \pi_{m}^{b}-\frac{1}{2} \pi^{ab} \pi\right)+h^{\frac{1}{2}}\left(\xi^{{\perp}\mid ab}-h^{ab} \xi^{{\perp}\mid m}{ }_{\mid m}\right) \\
		&+\left(\pi^{ab} \xi^{m}\right)_{\mid m}-\xi_{\mid m}^{a} \pi^{m b}-\xi^{b}{ }_{\mid m} \pi^{a m},
	\end{aligned}
\end{equation}
where $\mathcal{L}_{\xi} h_{ab}$ is the Lie derivative
\begin{align}
	\mathcal{L}_{\xi} h_{ab}=\xi_{a \mid b}+\xi_{b \mid a}.
\end{align}
\subsection{Preservation of falloff and gauge conditions}
From the preservation of the falloff conditions \eqref{falloff}-\eqref{falloff3} we obtain the demands
\begin{align}\label{demandhab}
	\delta h_{r r}=O\left(r^{-1}\right), \quad \delta h_{r A}=O(1), \quad \delta h_{A B}=O(r)
\end{align}
and we now want to find such $\xi^{\perp},\xi^a$ that the change of $h_{ab}$ defined in \eqref{transformationhab} obeys these demands.

Using the expansion of Christoffel symbols associated with $h_{ab}$
\begin{align}\label{christoffel}
	\Gamma^r_{AB}&=-r\bar\gamma_{AB}+O(1)\\
	\Gamma^A_{BC}&=\bar \Gamma^A_{BC}+O(r^{-1})\\
	\Gamma^r_{rA}&=\frac{1}{2r}\left(\partial_A \bar h_{rr}+\psi_A\right)+O(r^{-2})\\
	\Gamma^{r}_{rr}&=-\frac{\bar h_{rr}}{2r^2}+O(r^{-3})\\
	\Gamma^A_{rB}&=\frac{1}{r}\delta^A_B+O(r^{-2})\\
	\Gamma^A_{rr}&=-\frac{\bar\gamma^{AB}\partial_B \bar h_{rr}}{2r^3}+O(r^{-3}),
\end{align} 
we find the following transformation of $h_{rr}$
\begin{equation}
	\delta h_{rr}=\frac{2\xi^{\perp}}{\sqrt{\gamma}r^2}\left(\bar \pi_{rr}-\frac 12 \bar h_{rr}\bar \pi\right)+\xi^A\partial_A \frac{\bar h_{rr}}{r}+2\partial_r\xi^r-2\psi_A\partial_r \xi^A + \text{(subleading)},
\end{equation}
where
\begin{align}
    \bar \pi = \bar \pi^{rr}+\gamma_{AB}\bar \pi^{AB}.
\end{align}
Comparing this with \eqref{demandhab} we find that a large $r$ expansion of $\xi^{\perp},\xi^a$ has to be of the form
\begin{align}
	\xi^{\perp} = rb+F+O(r^{-1}),\quad \xi^A=Y^A+\frac 1r I^A+O(r^{-2}),\quad \quad \xi^r=W+\frac 1r \xi^r_1+O(r^{-2}),
\end{align}
where all terms in this expansion are functions on the sphere.
 
The transformation of $h_{rA}$
\begin{align}\label{transformhrA}
	\delta h_{rA}=\frac{2b}{\sqrt{\gamma}}\bar \pi_{rA}-\frac 12Y^BD_B \psi_A+\partial_AW-\frac 12\psi_BD_AY^B_0-\gamma_{AB}I^B + \text{(subleading)}
\end{align}
does not give any further restrictions on $\xi^{\perp},\xi^a$.

To fulfill the demand that $\delta h_{AB}$ 
\begin{align}\label{trafohAB}
	\delta h_{AB}=r^2D_{(A}Y_{B)}+r\Big(&2W\gamma_{AB}+Y^CD_C C_{AB}+\psi_{(B}\gamma_{A)C}Y^C+2C_{C(B}D_{A)}Y^C\nonumber\\
	&+2\gamma_{C(B}D_{A)}I^C
	+\frac{2b}{\sqrt{\gamma}}\left(\bar \pi_{AB}-\frac 12 \gamma_{AB}\bar \pi\right)
\end{align}
is of order $O(r)$ we have to assume that $Y^B$ are the Killing vectors on the 2-sphere
\begin{align}
	D_{(A}Y_{B)}=0.
\end{align}

So far there are no restrictions on $b$ but it is fixed from the preservation of the asymptotic form of momenta which demands
\begin{align}\label{momentademand}
	\delta \pi^{rr}=O(1),\quad \delta \pi^{rA}=O(r^{-1}),\quad \delta \pi^{AB}=O(r^{-2}).
\end{align}
One can check that there are no new restrictions coming from the first two conditions in \eqref{momentademand}. The third one, however, does lead to a new restriction and reads
\begin{align}\label{conditionb}
	\delta\pi^{AB}&=r^2\sqrt{\gamma}\left(\nabla^A\nabla^B\xi^{\perp}-h^{AB}\nabla^i\nabla_i\xi^{\perp}\right)+\text{(subleading)}\nonumber\\
	&=\frac{\sqrt{\gamma}}{r^2}\left(\gamma^{AC}\gamma^{AD}\nabla_C\nabla_D\xi^{\perp}-\gamma^{AB}\gamma^{CD}\nabla_C\nabla_D\xi^{\perp}\right)+\text{(subleading)}.
\end{align}
Now consider (we use the notation $\nabla_A=D_A+\text{(subleading)}$)
\begin{align}
	\nabla_C\nabla_D\xi^{\perp}&=r\partial_C\partial_Db-\Gamma^r_{CD}\partial_r(rb)-r\Gamma_{CD}^A\partial_Ab+O(1)\nonumber\\
	&=rD_CD_B b+r\gamma_{CD}b,
\end{align}
where we used \eqref{christoffel} and plugging this back into the previous equation we obtain
\begin{align}
	\delta \pi^{AB}=\frac 1r\left(D^AD^Bb-\gamma^{AB}D^2b-\gamma^{AB}b\right)+O(r^{-2}).
\end{align}
Vanishing of the leading order therefore imposes the condition
\begin{align}
	D_AD_Bb-\gamma_{AB}D^2b-\gamma_{AB}b=0 \Rightarrow D^2b=-2b
\end{align}
and therefore we find that $b$ has to fulfill the condition
\begin{align}
	D_AD_Bb+\gamma_{AB}b=0\,,
\end{align}
whose only solution is $b$ being a linear combination of three $l=1$ harmonics with constant coefficient.

Now we consider the preservation of the gauge condition $\bar h_{rA} = 0 $. From \eqref{transformhrA} one can directly see that in order to have $\delta \bar h_{rA}=0$ we need to assume
\begin{align}\label{IA}
	I^A=D^A W+\frac{2b}{\sqrt{\gamma}}\bar \pi^{rA},
\end{align}
which means that the preservation of the gauge choice 
\begin{align}
	\bar h_{rA}=-\frac 12\psi_A=0,
\end{align}
which we are going to adapt from now on, determines the subleading term of $\xi^A$. 

To summarize, $F$ and $W$ are not constrained by the boundary conditions and are associated with angle-dependent translations, temporal and spatial ones, respectively. $Y^A$ are the three Killing vectors on the sphere parametrizing rotations and $b$ contains only $l=1$ harmonics and parametrizes three boosts. Except for the assumptions of parity  on $W$ and $F$ we have therefore reproduced the asymptotic symmetry transformations given in \cite{Henneaux:2018cst} and the corresponding surface terms are therefore identical with \eqref{surfaceterms}, which has been derived for $F,W$ having no definite parity. Integrability of these surface terms demands that $F$ is of the form $F=-\frac 12 \bar h b+T(x^A)$, where $T(x^A)$ is a general function on the sphere. Any function of $\bar \lambda$ could be added to $F$ without spoiling integrability, which introduces an ambiguity in the expression of the charges. We choose for $F$ the form \eqref{defF}.

In \cite{Henneaux:2018cst} it is shown that transformations with $W$=odd, $T$=even form an algebra isomorphic to the BMS algebra found at null infinity. The odd and even functions combine to give the single function parametrizing supertranslations at null infinity. Since we do not involve any  parity conditions it appears that the resulting asymptotic symmetry is larger than the BMS symmetry, as long as the associated charges are finite. Before proceeding with the discussion of the charges, however, we turn to the aforementioned determinant condition.

\subsection{Determinant condition and symmetry reduction}\label{subsec:detcond}
In the previous subsection we have found that a large group of supertranslations and Lorentz transformations preserve the falloff conditions \eqref{falloff}-\eqref{falloff1}. We are now going to demonstrate that additionally imposing the determinant condition \eqref{detcond} breaks the invariance under all spatial translations, including the Poincar\'e ones.

{The determinant condition implies
\begin{align}\label{Ctraceless}
	\gamma^{AB}C_{AB}=0
\end{align}
and transformations preserving this condition must fulfill
\begin{align}
	\delta(\gamma^{AB}C_{AB})= \gamma^{AB}\delta \bar h_{AB}=0,
\end{align}
where we have used that $\delta \gamma_{AB}=0$ and $C_{AB}=\bar h_{AB}$.
Substituting \eqref{trafohAB} we find
\begin{align}\label{tracehAB}
	\gamma^{AB}\delta \bar h_{AB}=2(D^2+2)W+\frac{4}{\sqrt{\gamma}}D_A(b\bar \pi^{rA})-\frac{2b}{\sqrt{\gamma}}\bar \pi^{rr}=0
\end{align}
and using the expressions for the momenta \eqref{pirr} and \eqref{pirA} we obtain
\begin{equation}
	2(D^2+2)W=-2D^A(bD_A(2f-M))+b(4\mathcal M+D^2f-\gamma^{AB}\partial_u D_{AB}).
\end{equation}
This equation has no solution for $W$ since the RHS is in general non-vanishing and inevitably contains $l=1$ harmonics\footnote{In fact $b$ purely consists of $l=1$ harmonics and $(2f-\mathcal M)$ of $l=0,1$ harmonics.} which can not be produced by the LHS.}

We can therefore conclude that spatial translations, by which we mean standard Poincar\'e ones and supertranslations, do not preserve the condition \eqref{Ctraceless}, which shows that imposing the determinant condition excludes spatial translations from the group of asymptotic symmetries. 

This result is in contrast with the situation at null infinity.
The transformation of $C_{AB}$ there is given by (see for instance section 2.2 in \cite{Barnich:2011mi} for details concerning the asymptotic symmetries at null infinity)
\begin{align}
    -\delta C_{A B}=\left[f \partial_{u}+\mathcal{L}_{R}-\frac{1}{2} D_A R^A\right] C_{A B}-2 D_{A} D_{B} f+D_AD^A f \gamma_{A B},
\end{align}
where $R^A=Y^A-D^A b$ and $f=T+\frac 12 uD_AR^A$ parametrize Lorentz transformations and supertranslations which leave the asymptotic form of the Bondi metric at null infinity invariant. One can recognize from this expression that the condition $\gamma^{AB}C_{AB}=0$ is automatically preserved, since $\gamma^{AB}\delta C_{AB}=0$ under this condition. We therefore find that the symmetry algebra at spatial infinity is not obtained as the continuous limit $u \rightarrow -\infty$ of the one at null infinity. This is also reflected by the fact that the asymptotic symmetry transformations at null infinity themselves are in fact divergent in that limit, as is exemplified by the $u$-component of these transformations, $\xi^u=f$.

\subsection{The charges}

In the previous subsections we found that the asymptotic symmetry appears to be larger than the BMS symmetry at null infinity because there are two arbitrary functions on the sphere parametrizing translations. In the treatment of Henneaux and Troessaert \cite{Henneaux:2018cst} the charges associated with transformations outside of BMS are vanishing due to the imposed parity conditions and these transformations are therefore pure gauge. {Here we show that in our treatment the charges associated with all modes of $T$ and $W$ are in fact non-vanishing. We will also check that the boost and rotational charges are all non-vanishing.}
To obtain the form of the charges in terms of the spacetime metric functions we can use the expression \eqref{charge1} because it has been derived for general $T,W$, which corresponds to the case that we are considering.  We then only need to substitute the expressions for $h_{ab}$ and $\pi^{ab}$ in terms of \eqref{asympflat}, which we have already obtained from the 3+1 decomposition. From \eqref{charge1} the charges associated with translations are given as
\begin{align}\label{chargetranslation}
	\mathcal{B}_{T,W}=\oint d^{2} x\{2 T \sqrt{\gamma} \bar h_{rr}+ 2 W \left(\bar{\pi}^{r r}-\bar{\pi}_{A}^{A}\right)\},
\end{align}
the boost charges are
\begin{equation}\label{chargeboost}
	\mathcal{B}_{b}=\oint d^{2} x\left[b \sqrt{\gamma}\left(2 k^{(2)}+\bar{k}^{2}+\bar{k}_{B}^{A} \bar{k}_{A}^{B}-6 \overline{\lambda k}\right)+b \frac{2}{\sqrt{\gamma}} \gamma_{A B} \bar{\pi}^{r A} \bar{\pi}^{r B}\right]
\end{equation}
and finally the rotational charges
\begin{equation}\label{chargerot}
	\mathcal{B}_{Y}=\oint d^{2} x Y^{A}\left(4 \bar{k}_{A B} \bar{\pi}^{r B}-4 \bar{\lambda} \gamma_{A B} \bar{\pi}^{r B}+2 \gamma_{A B} \pi^{(2) r B}\right),
\end{equation}
where $k^{(2)}$ is defined via the expansion

\begin{align}
	K_{B}^{A} &=h^{A C} K_{B C}=-\frac{1}{r} \delta_{B}^{A}+\frac{1}{r^{2}} \bar{k}_{B}^{A}+\frac{1}{r^{3}} k_{B}^{(2) A}+O\left(r^{-3}\right)\label{expansionK} \\
	K_{A B} &=\frac{1}{2 \lambda}\left(-\partial_{r} h_{A B}+\nabla_{A} h_{r B}+\nabla_{B} h_{r A}\right)\label{defK} \\
	\lambda &=\frac{1}{\sqrt{h^{r r}}}\label{deflambda}.
\end{align}

Substituting the expressions for the momenta \eqref{pirr} and \eqref{piBA} into \eqref{chargetranslation} yields for the translational charge 
\begin{align}
	\mathcal{B}_{W, T}=\oint d^{2} x \sqrt{\gamma} \{T 2\bar h_{rr}+2W(\mathcal M+(D^2+2)f -\frac 12 \gamma^{AB}\partial_u D_{AB}-2A)\}
\end{align}
and using the condition \eqref{condfM} we can eliminate $f$ and $A$ from this expression and obtain
\begin{align}\label{transcharge}
	\mathcal{B}_{W, T}=\oint d^{2} x \sqrt{\gamma} \{T 2\bar h_{rr}+W((D^2+4)\mathcal M-\gamma^{AB}\partial_u D_{AB})\}.
\end{align}
{To investigate which modes of $T, W$ lead to finite charges we expand them and the metric functions in spherical harmonics and use their orthonormality and the fact that any spherical harmonic by itself vanishes when integrated over the sphere. This implies that the only finite terms are the ones where each factor has a contribution from the same mode. If $\bar h_{rr}$, for instance, was a constant then the first term in the above charge would only have a non-vanishing contribution from the zero mode of $T$. 
We can therefore see that there are finite contributions from all modes of $T$ and $W$ present, since only the combination $f-2\mathcal M$ is constrained to $l=0$ and $l=1$ modes but $\mathcal M$ itself contains in general contributions from all modes and so does $\bar h_{rr}$.}

As a sanity check we consider the Vaidya spacetime defined by 
\begin{align}
    ds^2=-\left(1-\frac{2M(u)}{r}\right)du^2-2dudr+r^2\gamma_{AB}dx^Adx^B,
\end{align}
which is a special case of \eqref{asympflat} with $M=M(u)$ and all subleading components vanishing. The ADM mass of this spacetime is obtained by setting $T=1, W=0$ in \eqref{transcharge} in which case we obtain
\begin{align}
    M_{\text{ADM}}=\lim_{u\rightarrow -\infty}16\pi M(u),
\end{align}
which agrees with the expression given in \cite{Poisson:2009pwt}(chapter 4.3.5 therein), up to the normalization factor $16\pi$. In particular this also implies that we obtain the correct expression for the ADM mass for the Schwarzschild spacetime, which is obtained by further specializing to the case $M=const.$.

{Next we are going to write the rotation and boost charges in terms of the spacetime metric functions and in doing so we will specialize to the case $f=const.$ and $g=const.$, which will make the otherwise lengthy expressions much more compact. As we are going to argue in a moment this choice has no impact on which modes of the charges are vanishing or not.
For $f=const., g=const.$ we find for the subleading contribution to $\pi^{rA}$
\begin{align}
	\pi^{(2)rA}=&\frac{\sqrt{\gamma}}{4}\Big(D^A\mathcal M\gamma^{AB}C_{AB}-4C^{AB}D_B \mathcal M+3F^A\nonumber\\
	&-2D^Ag_{ur}^{(2)}+2D^A \bar g_{uu}-2D^A\mathcal M\partial_u \bar g_{uu}-4MD^A \bar g_{ur}\nonumber\\
	&+16 MD^A M-2\bar g_{ur}D^A \bar g_{ur}\Big)\label{pi2rA}
\end{align}
and together with 
\begin{equation}
	4\bar k_{AB}\bar \pi^{rB}-4\bar \lambda = -2C_{AB}D^B\mathcal M,
\end{equation}
where we have used the definition \eqref{defkAB}, we find that the rotational charge \eqref{chargerot} 
\begin{equation}\label{chargerot}
	\mathcal{B}_{Y}=\oint d^{2} x Y^{A}\left(-2C_{AB}D^B\mathcal M+2 \gamma_{A B} \pi^{(2) r B}\right)
\end{equation}
is indeed non-vanishing for all modes of $Y^A$ since $F^A$ and $\mathcal M$ contain in general contributions from all modes of spherical harmonics. Allowing for general $f$ and $g$ would not change this conclusion in particular because there is no condition that would relate $F_A$ to either of these functions.
To obtain the expression for the boost charge \eqref{chargeboost} we first evaluate \eqref{defK} and \eqref{deflambda} using \eqref{identify} and find
\begin{align}
	K_{AB}=\frac{1}{2\lambda}\left(-2r\gamma_{AB}-C_{AB}-\frac{1}{2r}D_{(A}F_{B)}\right)+O(r^{-4})
\end{align}
and
\begin{align}
	\frac{1}{\lambda}=1-\frac{\bar h_{rr}}{r}+\frac{L}{r^2}+O(r^{-3}),
\end{align}
where
\begin{align}
	L=-4\bar g_{uu}+12(\mathcal M\bar g_{ur}+M^2)+8g_{ur}^{(2)}.
\end{align}
To calculate $k^{(2)}$ from \eqref{expansionK} we also need
\begin{align}
	h^{AB}=\frac{1}{r^2}\gamma^{AB}-\frac{1}{r^{3}} C^{AB}+\frac{1}{r^4}\left(C^A_DC^{DB}-D^{AB}\right)+O(r^{-5})
\end{align}
and find
\begin{equation}
	k^{(2)}=2\gamma_{A B}D^{AB}-C_{AB}C^{AB}-4L-D_AF^A
\end{equation}
and finally obtain for the boost charge
\begin{align}
	\mathcal{B}_{b}=\oint d^{2} x \sqrt{\gamma} b&\Big(D_A\mathcal MD^A\mathcal M+2\gamma_{A B}D^{AB}-4L-D_AF^A\nonumber\\
	&+\frac 14 (\gamma_{A B}C^{AB})^2-\frac 34 C_{AB}C^{AB}-\frac 32 \bar h_{rr}^2-\frac 52 \gamma_{A B}C^{AB}\bar h_{rr}\Big).
\end{align}
Again we can see that the charge is finite for all modes of $b$ since $F^A, \mathcal M$ and $\bar h_{rr}$ contain contributions from all modes.}


\subsection{Discussion of the asymptotic symmetries}
From our discussion of the charges it has become clear that the asymptotic symmetry we find at spatial infinity is larger than the BMS symmetry. The crucial difference to the results of \cite{Henneaux:2018cst} is that the charges associated with even $W$ and odd $T$ are non-vanishing. 

A larger-than-BMS asymptotic symmetry at spatial infinity has been found previously by Ashtekar and Hansen \cite{Ashtekar:1978zz}. The Spi algebra they find has the same structure as BMS, namely a semi-direct product of the abelian ideal of supertranslations and the Lorentz algebra. The difference lies in the size of the supertranslation ideal, which for the BMS algebra corresponds to functions on the 2-sphere whereas for Spi it corresponds to functions on the three-dimensional hyperboloid. 

Also Troessaert \cite{Troessaert:2017jcm} finds an asymptotic symmetry at spatial infinity which is larger than BMS, but smaller than the Spi algebra. By additionally assuming that the spacetime metric considered therein has to be asymptotically flat not only at spatial infinity but also at null infinity conditions on the metric functions are found which reduce the algebra to one that is isomorphic to the BMS algebra. This algebra is then in turn shown by \cite{Henneaux:2018cst} to be isomorphic to the algebra \eqref{algatspatial} with odd $W$ and even $T$.
For arbitrary $W$ and $T$ the algebra \eqref{algatspatial} is in fact also isomorphic to the one found by \cite{Troessaert:2017jcm}, before cutting it down to BMS. We recall some details of this construction in Appendix B.

{To summarize, since we do not impose parity conditions} we find an asymptotic symmetry at spatial infinity that is larger than the BMS algebra but smaller than the Spi algebra. Our result would therefore suggest that the tension arising from the presence of different asymptotic symmetries at spatial infinity and null infinity still remains.

\section{Conclusions}\label{sec:conclusions}
We have analyzed the asymptotic symmetries of asymptotically flat spacetimes in the Hamiltonian formulation of GR. In contrast to previous treatments we have expressed the asymptotic expansion of the  spatial metric and conjugate momenta in terms of a Bondi-type  spacetime metric \eqref{asympflat} using a 3+1 decomposition. {An important insight of this procedure was that the falloff conditions on the momenta translate to conditions on retarded time derivatives of spacetime metric functions which describe the rate of gravitational radiation. If only spacetimes are allowed that radiate a finite amount of energy the falloff conditions are automatically satisfied and we furthermore find that then also the kinetic term in the action is finite without having to introduce parity conditions.}

{As a consequence we found that an enlarged sector of supertranslations is present in the asymptotic symmetry, which is parametrized by two arbitrary functions on the sphere. The associated charges were found to be finite for every mode of these two functions.}
Our results therefore suggest that the supertranslation sector at spatial infinity is larger than the one of the BMS algebra, which is parametrized by a single arbitrary function on the sphere. 

A result that remains to be understood better is that spatial translations do not preserve the Bondi determinant condition in our treatment. Why is it that at null infinity this condition is fulfilled automatically whereas at spatial infinity it turns out to be too rigid? Another intriguing question in this context is whether the supertranslation sector at null infinity can be enlarged by {relaxing the determinant condition in an appropriate way}. Possibly this enlarged algebra at null infinity is isomorphic to the one we found at spatial infinity? We hope to address these questions in the future.

\section*{Acknowledgment}
This work was supported by funds provided by the National Science Center, projects number 2017/27/B/ST2/01902 and 2019/33/B/ST2/00050.

\appendix
\section{3+1 decomposition}

Following  \cite{Poisson:2009pwt} we briefly recall the main features of the 3+1 decomposition. Such a decomposition is obtained by introducing a foliation of spacetime into spacelike hypersurfaces $\Sigma_t$, defined by $t=const.$, where the only condition on $t$ is that the unit normal $n_{\alpha} \propto \partial_{\alpha}t$ has to be a future directed timelike vector field. We will show in the next section how different choices of $t$ affect the conjugate momenta and the parity conditions. Further, one introduces a time-evolution vector field $t^{\alpha}$ to define the direction of time derivatives. The defining condition is $t^{\alpha}\partial_{\alpha}t=1$, which allows to interpret the directional derivative $t^{\alpha}\partial_{\alpha}$ as $\partial_t$ and thus ensures that the direction of time derivatives is compatible with the meaning of time provided by $t$. Using $t^{\alpha}$ one can define $\dot h_{ab} = \pounds_t h_{ab}$ and conjugate momenta
\begin{equation}\label{defmomenta}
	\pi^{ab}=\frac{\partial }{\partial \dot h_{ab}}\mathcal L_G,
\end{equation}
where $h_{ab}$ is the induced metric on $\Sigma_t$ and $\mathcal L_G$ is the gravitational Lagrangian.
The vector field $t^{\alpha}$ is usually decomposed into parts tangential and orthogonal to the spatial hypersurfaces
\begin{equation}\label{timeflow}
	t^{\alpha}=Nn^{\alpha}+N^ae^{\alpha}_a,
\end{equation}
where $e^{\alpha}_a=\frac{\partial x^{\alpha}}{\partial y^a}$ are the tangent vectors on $\Sigma_t$, $x^{\alpha}$ are coordinates in the full spacetime and $y^a$ are coordinates intrinsic to $\Sigma_t$. $N$ is referred to as lapse and $N^i$ as shift.
Using this decomposition of $t^{\alpha}$ one can write
\begin{align}
	dx^a&=\frac{\partial x^{\alpha}}{\partial t}dt+\frac{\partial x^{\alpha}}{\partial y^{\alpha}}dy^a=t^{\alpha}dt+e^{\alpha}_ady^a
\end{align}
and for the line element of a generic metric it follows
\begin{equation}\label{decomp}
	ds^2=g_{\alpha\beta}dx^{\alpha}dx^{\beta}=-N^2dt^2+h_{ab}(dy^a+N^a dt)(dy^b+N^b dt),
\end{equation}
where the definition of the induced metric 
\begin{align}\label{definduced}
	h_{ab}=g_{\alpha\beta}e^{\alpha}_ae^{\beta}_b
\end{align}
has been used. Note, that asymptotic flatness demands for lapse and shift to behave asymptotically as \cite{Regge:1974zd},\cite{Henneaux:2018hdj}
\begin{align}\label{lapseasmyp}
	N=1+O(1/r),\; N^r = O(1/r),\; N^A=O(1/r^2).
\end{align}
The conjugate momenta can be expressed as
\begin{equation}\label{momenta}
	\pi^{ab}=\sqrt{h}(K^{ab}-Kh^{ab})
\end{equation}
and we are going to use this expression to write the conjugate momenta in terms of $g_{\alpha\beta}$ and its derivatives.
The extrinsic curvature $K_{ab}$ is defined as
\begin{equation}\label{extrinsic}
	K_{ab} = n_{\alpha;\beta}e^{\alpha}_ae^{\beta}_b,
\end{equation}
where semicolon denotes the covariant derivative associated with $g_{\alpha\beta}$ and $K=h^{ab}K_{ab}$ is the trace of extrinsic curvature. Finally, $\dot h_{ab}$ can be written as
\begin{align}\label{velocity}
	\dot h_{ab} = \mathcal L_t h_{ab} &= 2NK_{ab}+N_{a|b}+N_{b|a}\\
	&=\frac{2N}{\sqrt{h}}\left(\pi_{ab}-\frac 12\pi h_{ab}\right)+N_{a|b}+N_{b|a},
\end{align}
where $\pi=h_{ab}\pi^{ab}$ and $N_{a|b}$ denotes the covariant derivative associated with $h_{ab}$.

\section{The algebra of asymptotic symmetries}

In this appendix we show that the algebra \eqref{algatspatial} is for arbitrary $W$ and $T$ isomorphic to the one found in \cite{Troessaert:2017jcm}, using arguments presented in the appendix of \cite{Henneaux:2018cst}. In
 \cite{Troessaert:2017jcm} the following asymptotic algebra is found
\begin{equation}\label{algtro}
	\left[\left(\mathcal Y_{1}, \omega'_{1}\right),\left(\mathcal Y_{2}, \omega'_{2}\right)\right]=\left(\left[\mathcal Y_{1}, \mathcal Y_{2}\right], \mathcal Y_{1}^{a} \partial_{a} \omega'_{2}-\frac{s}{2} \psi_{1} \omega'_{2}-(1 \leftrightarrow 2)\right)=\left(\widehat{\mathcal Y}, \widehat \omega '\right),
\end{equation}
where $x^a=(s,x^A)$ are coordinates on the unit hyperboloid, $\mathcal Y^a$ represents the Lorentz algebra, $\omega' = \sqrt{1-s^2}\omega$ and $\omega(x^a)$ parametrize a sub-set of Spi supertranslations \cite{Ashtekar:1978zz}. Full Spi supertranslations would be given by general functions $\omega(x^a)$ but, as \cite{Troessaert:2017jcm} explains, to remove divergences in the symplectic structure one has to demand
\begin{align}
	(D^aD_a + 3)\omega = 0.
\end{align}
The general solution of this equation is shown to be of the form 
\begin{equation}
\begin{gathered}
\omega=\frac{1}{\sqrt{1-s^{2}}}\left(\widehat{\omega}^{\text {even }}+\widehat{\omega}^{o d d}\right) \\
\widehat{\omega}^{\text {even }}=\sum_{l, m} \widehat{\omega}_{l m}^{V} V_{l}(s) Y_{l m}^{0}\left(x^{A}\right), \quad \widehat{\omega}^{\text {odd }}=\sum_{l, m} \widehat{\omega}_{l m}^{W} W_{l}(s) Y_{l m}^{0}\left(x^{A}\right),
\end{gathered}
\end{equation}
where odd and even refers to the combination of time reversal $s\rightarrow -s$ and antipodal mapping $x^A\rightarrow -x^A$ and $V_l(s), W_l(s)$ are defined in terms of Legendre polynomials and Legendre functions of the second kind.

Rotations are parametrized by Killing vectors on the 2-sphere $\mathcal Y^A_R(x^A)$
\begin{align}
	\mathcal Y^s=0,\quad \mathcal Y^A=\mathcal Y^A_R
\end{align}  
and boosts by $\psi(x^A)$ such that $D^2 \psi+2\psi=0$
\begin{equation}
	\mathcal Y^{s}=-\frac{1}{2}\left(1-s^{2}\right) \psi, \quad \mathcal{Y}^{A}=-\frac{1}{2} s \gamma^{A B} \partial_{B} \psi.
\end{equation}

One can check that the Lorentz algebras in \eqref{algtro} and \eqref{algatspatial} are isomorphic under the identification $\mathcal Y^A_R=Y^A$ and $\psi=2b$. The action of the Lorentz algebra on $\omega'$ in \eqref{algtro} can then be written as
\begin{equation}\label{actionLor}
	\widehat \omega'=Y^{A}_1 \partial_{A} \omega'_2-s b_1 \omega'_2-s \partial^{A} b_1 \partial_{A} \omega'_2-\left(1-s^{2}\right) b_1 \partial_{s} \omega'_2-(1 \leftrightarrow 2).
\end{equation}
The connection with the ADM description in \eqref{algatspatial} can be made by defining $W,T$ as initial conditions at $s=0$
\begin{equation}\label{defWT}
	\left.\omega\right|_{s=0}=\left.\omega'\right|_{s=0}=W\left(x^{A}\right),\left.\quad \partial_{s} \omega\right|_{s=0}=\left.\partial_{s} \omega'\right|_{s=0}=T\left(x^{A}\right).
\end{equation}
One can check from the definitions of $V_l(s)$ and $W_l(s)$ that $\left.\omega\right|_{s=0}$ and $\left.\partial_{s} \omega\right|_{s=0}$  contain contributions from all modes of spherical harmonics and that therefore $W$ and $T$ such defined are arbitrary functions on the sphere.
Substituting these definitions in \eqref{actionLor} one obtains
\begin{align}\label{W_hat}
	\widehat W=Y_1^A\partial_A W_2-b_1T_2-(1 \leftrightarrow 2).
\end{align}
Acting with the $s$ derivative on \eqref{actionLor} yields
\begin{align}\label{sderiv}
	\partial_s \widehat \omega'=&Y_1^A\partial_A\partial_s\omega'_2-b_1 \omega'_2-sb_1\partial_s\omega'_2- \partial^{A} b_1 \partial_{A} \omega'_2-s\partial^Ab_1\partial_A\omega_2'\nonumber\\
	&+2sb_1\partial_s\omega_2'-(1-s^2)b_1\partial_s^2\omega_2'-(1 \leftrightarrow 2).
\end{align}
The expression for $\left.\partial_s^2 \omega'\right|_{s=0}$ can be obtained from
\begin{equation}
	\left(\mathcal{D}_{a} \mathcal{D}^{a}+3\right) \omega=-\left(1-s^{2}\right)^{2} \partial_{s}^{2} \omega+\left(1-s^{2}\right) D^2 \omega+3 \omega=0
\end{equation}
and first one has
\begin{align}
	\left.\partial_s^2 \omega\right|_{s=0} = \left.D^2\omega\right|_{s=0} + \left.3\omega\right|_{s=0}
\end{align}
and together with
\begin{align}
	\left.\partial_s^2 \omega'\right|_{s=0} = \left.-\omega\right|_{s=0} + \left.\partial_s^2 \omega\right|_{s=0} 
\end{align}
one finally has
\begin{align}
	\left.\partial_s^2 \omega'\right|_{s=0} = \left. 2\omega\right|_{s=0} + \left .D^2\omega\right|_{s=0}.
\end{align}
Substituting this and the above definitions in \eqref{sderiv} and evaluating at $s=0$ one finally obtains
\begin{equation}\label{T_hat}
	\widehat T = Y_{1}^{A} \partial_{A} T_{2}-3 b_{1} W_{2}-\partial_{A} b_{1} D^{A} W_{2}-b_{1}D^2 W_{2}-(1 \leftrightarrow 2).
\end{equation}
The expressions \eqref{W_hat} and \eqref{T_hat} agree with the corresponding ones in \eqref{algatspatial}.


\begin{thebibliography}{99}

	\bibitem{Henneaux:2018cst}
M.~Henneaux and C.~Troessaert,
``BMS Group at Spatial Infinity: the Hamiltonian (ADM) approach,''
JHEP \textbf{03} (2018), 147
[arXiv:1801.03718 [gr-qc]].

\bibitem{Freidel:2020xyx}
L.~Freidel, M.~Geiller and D.~Pranzetti,
``Edge modes of gravity. Part I. Corner potentials and charges,''
JHEP \textbf{11} (2020), 026
[arXiv:2006.12527 [hep-th]].

	\bibitem{Regge:1974zd}
T.~Regge and C.~Teitelboim,
``Role of Surface Integrals in the Hamiltonian Formulation of General Relativity,''
Annals Phys. \textbf{88} (1974), 286

\bibitem{Bondi:1962px}
  H.~Bondi, M.~G.~J.~van der Burg and A.~W.~K.~Metzner,
  ``Gravitational waves in general relativity. 7. Waves from axisymmetric isolated systems,''
  Proc.\ Roy.\ Soc.\ Lond.\ A {\bf 269} (1962) 21.
  
\bibitem{Sachs:1962wk}
R.~K.~Sachs,
``Gravitational waves in general relativity. 8. Waves in asymptotically flat space-times,''
Proc. Roy. Soc. Lond. A \textbf{270} (1962), 103-126

\bibitem{Sachs:1962zza}
R.~Sachs,
``Asymptotic symmetries in gravitational theory,''
Phys.\ Rev.\  {\bf 128}, 2851 (1962).

\bibitem{Strominger:2017zoo}
A.~Strominger,
``Lectures on the Infrared Structure of Gravity and Gauge Theory,''
[arXiv:1703.05448 [hep-th]].




\bibitem{Troessaert:2017jcm}
C.~Troessaert,
Class. Quant. Grav. \textbf{35} (2018) no.7, 074003
doi:10.1088/1361-6382/aaae22
[arXiv:1704.06223 [hep-th]].

\bibitem{Brown:1986nw}
J.~D.~Brown and M.~Henneaux,
Commun. Math. Phys. \textbf{104} (1986), 207-226
doi:10.1007/BF01211590

\bibitem{Teitelboim:1972vw}
C.~Teitelboim,
Annals Phys. \textbf{79} (1973), 542-557
doi:10.1016/0003-4916(73)90096-1

\bibitem{Ciambelli:2021nmv}
L.~Ciambelli, R.~G.~Leigh and P.~C.~Pai,
``Embeddings and Integrable Charges for Extended Corner Symmetry,''
[arXiv:2111.13181 [hep-th]].

\bibitem{Freidel:2021dxw}
L.~Freidel,
``A canonical bracket for open gravitational system,''
[arXiv:2111.14747 [hep-th]].

\bibitem{Arnowitt:1962hi}
R.~L.~Arnowitt, S.~Deser and C.~W.~Misner,
``The Dynamics of general relativity,''
Gen. Rel. Grav. \textbf{40} (2008), 1997-2027
doi:10.1007/s10714-008-0661-1
[arXiv:gr-qc/0405109 [gr-qc]].

\bibitem{Flanagan:2015pxa}
\'E.~\'E.~Flanagan and D.~A.~Nichols,
``Conserved charges of the extended Bondi-Metzner-Sachs algebra,''
Phys. Rev. D \textbf{95} (2017) no.4, 044002
doi:10.1103/PhysRevD.95.044002
[arXiv:1510.03386 [hep-th]].

\bibitem{Barnich:2009se}
G.~Barnich and C.~Troessaert,
``Symmetries of asymptotically flat 4 dimensional spacetimes at null infinity revisited,''
Phys. Rev. Lett. \textbf{105} (2010), 111103
doi:10.1103/PhysRevLett.105.111103
[arXiv:0909.2617 [gr-qc]].

	\bibitem{Poisson:2009pwt}
E.~Poisson,
``A Relativist's Toolkit: The Mathematics of Black-Hole Mechanics,'' Cambridge University Press, Cambridge, 2004.
doi:10.1017/CBO9780511606601

\bibitem{Barnich:2011mi}
G.~Barnich and C.~Troessaert,
``BMS charge algebra,''
JHEP \textbf{12} (2011), 105
doi:10.1007/JHEP12(2011)105
[arXiv:1106.0213 [hep-th]].

	
\bibitem{Henneaux:2018hdj}
M.~Henneaux and C.~Troessaert,
``Hamiltonian structure and asymptotic symmetries of the Einstein-Maxwell system at spatial infinity,''
JHEP \textbf{07} (2018), 171
doi:10.1007/JHEP07(2018)171

	\bibitem{Barnich:2010eb}
G.~Barnich and C.~Troessaert,
``Aspects of the BMS/CFT correspondence,''
JHEP \textbf{05} (2010), 062
doi:10.1007/JHEP05(2010)062
[arXiv:1001.1541 [hep-th]].




\bibitem{Compere:2018aar}
G.~Comp\`ere and A.~Fiorucci,
``Advanced Lectures on General Relativity,''
[arXiv:1801.07064 [hep-th]].

\bibitem{Ashtekar:1978zz}
A.~Ashtekar and R.~O.~Hansen,
J. Math. Phys. \textbf{19} (1978), 1542-1566
doi:10.1063/1.523863

\end{thebibliography}
\end{document}